\documentclass[aip,jcp,reprint]{revtex4-2}
\usepackage{chemformula} 
\usepackage[T1]{fontenc} 
\usepackage{comment}
\def\lang{\langle}
\def\rang{\rangle}

\begin{document}

\title{Quantum Many-Body Linear Algebra, Hamiltonian Moments, and a Coupled Cluster Inspired Framework}

\author{Yuhang Ai}
\email{yai@caltech.edu}
\affiliation{Division of Chemistry and Chemical Engineering, California Institute of Technology,
Pasadena, California 91125, USA}
\author{Huanchen Zhai}
\affiliation{Division of Chemistry and Chemical Engineering, California Institute of Technology,
Pasadena, California 91125, USA}
\author{Johannes T{\"o}lle}
\affiliation{Division of Chemistry and Chemical Engineering, California Institute of Technology,
Pasadena, California 91125, USA}
\author{Garnet Kin-Lic Chan}
\email{gkc1000@gmail.com}
\affiliation{Division of Chemistry and Chemical Engineering, California Institute of Technology,
Pasadena, California 91125, USA}

\begin{abstract}
        We propose a general strategy to develop quantum many-body approximations of primitives in linear algebra algorithms. As a practical example, we introduce a coupled-cluster inspired framework to produce approximate Hamiltonian moments, and demonstrate its application in various linear algebra algorithms for ground state estimation. Through numerical examples, we illustrate the difference between the ground-state energies arising from quantum many-body linear algebra  and those from the analogous many-body perturbation theory. Our results support the general idea of designing quantum many-body approximations outside of perturbation theory, providing a route to new algorithms and approximations.
\end{abstract}

\maketitle

\section{Introduction}

Hamiltonian moments $M_n=\lang \Phi|\hat H^n|\Phi\rang$ appear
in many linear algebra algorithms to determine the eigenvalues of $\hat H$, e.g. exact diagonalization and kernel polynomial approximations\cite{Saad1992,AlexanderWeisse2008}. The exact evaluation of $\lang \Phi|\hat H^n|\Phi\rang$, however, quickly becomes intractable for large problems.
Different approximations, for example, based on using a sparse approximation to $\hat{H}^n|\Phi\rangle$  (as featured in selected configuration interaction methods\cite{Booth2009,Tubman2016,Liu2016,Sharma2017}), or representing $\hat H$ and $|\Phi\rang$ by an ansatz (as used in tensor network algorithms\cite{White1992,ChanGKL2011,Cirac2021}) have been proposed, allowing for affordable approximate computation of the moments.

Alternatively, using Wick's theorem, expectations of operator products like $\lang \Phi|\hat H^n|\Phi\rang$ can be expanded into a set of operator diagrams representing the underlying contraction patterns\cite{Shavitt2009}.
The number of such contractions or diagrams, and the cost of the most expensive diagrams, grows exponentially with $n$ and system size\cite{Chan2024}, and thus approximations are still needed. 
Previously, diagrammatic approximations have been investigated in the context of many-body perturbation theory\cite{Shavitt2009} (MBPT). In addition to operator matrix elements, such diagrams contain energy denominators or Green's functions. A vast literature exists concerning the classification of these diagrams and their subsequent summation and resummation according to their topologies and physical content,  encompassing many-body approaches such as coupled-cluster (CC) theory\cite{Shavitt2009}, self-consistent Green's function methods\cite{Phillips2014,Golze2019}, and diagrammatic quantum Monte Carlo\cite{Prokofev1998,EmanuelGull2011}. 

In this Communication, we explore the possibility of building non-perturbative diagrammatic approximations to the expectations of operator products, and the Hamiltonian moments in particular. These diagrams do not contain energy denominators or Green's functions and are thus simpler than those appearing in MBPT. Once approximated, the moments can be used in a linear algebra algorithm, an approach we refer to as quantum many-body linear algebra. As a concrete realization, we describe the generation of Hamiltonian moments using diagrams classified by a coupled cluster-like framework. As we shall illustrate, this opens up new ways to build many-body approximations  that behave differently from those derived from MBPT.

\section{Linear Algebra with Moments and Quantum Many-Body Approximations \label{sec:la}}

Although linear algebra algorithms may involve multiple linear operators, for simplicity of discussion we focus on algorithms involving the single operator $\hat{H}$, where we aim to approximate its ground-state eigenvalue $E_0$. As discussed in the introduction, many such algorithms implicitly compute $E_0$ from the  moments of $\hat{H}$.
We will consider the following representative examples:
\begin{itemize}
    \item Power Method. Consider the operator $\hat{F} = \lambda - \hat{H}$, where $\lambda$ satisfies $\lambda >(E_0+E_\mathrm{max})/2$ ($E_0$, $E_\mathrm{max}$ being extremal eigenvalues of $\hat{H}$).
    We obtain $E_0$ in the limit
    \begin{align}
        E_0 = \lim_{k\to \infty} \frac{\langle \Phi |\hat{F}^k \hat{H} \hat{F}^k |\Phi\rangle}{\langle \Phi | \hat{F}^{2k} | \Phi\rangle} = \lim_{k\to \infty} E_k^\mathrm{P}
    \end{align}
    The numerator and denominator can be expressed in terms of $M_n$, and we need up to $M_{2k+1}$ ($M_{2k}$) to compute the numerator (denominator) for a given iteration number $k$. 

    \item Power Method with Chebyshev acceleration. A variant of the power method, the Chebyshev iteration\cite{Saad1992} builds the vectors $|\Phi^\mathrm{C}_k\rang = T_k((\hat H-c)/e)|\Phi\rang/T_k((\nu-c)/e)$, where $T_k(x)$ is the Chebyshev polynomial of degree $k$ of the first kind, $\nu$ is an initial estimate of $E_0$, and $(c,e)$ defines an ellipse in the complex plane with center $c$ and foci $c+e,c-e$ that covers all eigenvalues other than $E_0$. We obtain $E_0=\lim_{k\to\infty} \lang \Phi_k^\mathrm{C}|\hat H|\Phi_k^\mathrm{C}\rang/\lang\Phi_k^\mathrm{C}|\Phi_k^\mathrm{C}\rang = \lim_{k\to\infty} E_k^\mathrm{C} $. Again, the numerator and denominator can be expressed in terms of $M_n$ at each iteration. 

    \item Lanczos Diagonalization. Given an initial $|\Phi\rang$, the $k$-th iteration of Lanczos corresponds to diagonalizing the Hamiltonian in the Krylov subspace $\{|\Phi_{k}^\mathrm{K}\rang=\hat H^k|\Phi\rang\}$, where the matrix representation of $\hat{H}$ takes the form of a tridiagonal Jacobi matrix
    \begin{equation}
        \mathbf{H}_{k} = \begin{pmatrix}
            a_0 & b_1 & ~ & ~ & ~ & ~\\ 
            b_1 & a_1 & b_2 & ~ & ~ & ~ \\ 
            ~ & b_2 & a_2 & \ddots & ~ & ~\\ 
            ~ & ~ & \ddots & \ddots & b_{k-1} & ~ \\
            ~ & ~ & ~ & b_{k-1} & a_{k-1} & b_{k} \\ 
            ~ & ~ & ~ & ~ & b_{k} & a_{k} \\  
        \end{pmatrix}
        \label{eqn:krylov_ham}
    \end{equation}
    with the elements defined according to the standard three-term recurrence relation\cite{AlexanderWeisse2008}
    \begin{align}
        b_{k+1}|v_{k+1}\rang =|\tilde v_{k+1}\rang= \hat H|v_k\rang -a_k |v_k\rang -b_k |v_{k-1}\rangle \notag \\
    \end{align}
    starting from $|v_0\rang=|\tilde v_0\rang=|\Phi\rang$, and with $ a_k = \lang v_k|\hat H|v_k\rang$ and $b_k = \lang \tilde v_k|\tilde v_k\rang$. Again, these values can be written in terms of the moments, and we need up to $M_{2k+1}$ ($M_{2k}$) to compute $a_k$ ($b_k$) for a given iteration number $k$. 

\end{itemize}

All the above algorithms yield the exact ground-state energy $E_0$ when performed for sufficient iterations with exact moments, and for any finite $k$ yield a variational approximation to $E_0$. They are thus equivalent aside from the rate of convergence. There are also related methods based on moments that yield non-variational estimates of $E_0$, notably the connected moments expansion\cite{Cioslowski1987}, which we do not explore here.

The quantum many-body version of such algorithms is obtained when we use a many-body approximation to the Hamiltonian moments. We develop such many-body approximations using Wick's theorem to expand $M_n$ into a sum of products of terms. Different versions of Wick's theorem may be applied depending on the nature of $|\Phi\rangle$. For example, if $|\Phi\rang=|\phi\rang$ is a single determinant, we can use the single determinant version of Wick's theorem. If $|\Phi\rangle = \sum_{I}C_I |\phi_I\rang$ is a sum of determinants, we may express $M_n$ in terms of the transition moments $\langle \phi_I | \hat{H}^n | \phi_J\rangle$ where each is evaluated by Wick's theorem. Alternatively, one may apply the generalized Wick's theorem\cite{Kutzelnigg1997} to write $\langle \Phi | \hat{H}^n |\Phi\rangle$ as a sum of contributions involving the many-body cumulants. Regardless of the version of Wick's theorem, each term corresponds to a particular pattern of contraction of the fermionic operators in the $n$ copies of $\hat{H}$, that can be recorded as a diagram, and the approximation arises from keeping a subset of diagrams. A systematic truncation then yields the many-body approximation to the linear algebra.

\subsection{Termination criteria for approximate quantum many-body linear algebra}

The approximate moments do not satisfy all the properties of the exact moments, which has implications for the corresponding linear algebra algorithms. The convergence of standard  algorithms relies on the properties of the higher moments (i.e. for large $k$) which are precisely those that are likely to be least accurate within any approximation. This suggests that modified termination criteria should be used in quantum many-body linear algebra.

For example, the power method is guaranteed to converge to a fixed point when using the exact moments, but this is not so with arbitrary approximations to the higher moments. Similarly, the Lanczos algorithm assumes that the overlap matrix of Krylov vectors $S_{ij} = \lang \Phi|\hat H^i \hat H^j|\Phi\rang = M_{i+j}$ is  positive definite.
Hamburger's theorem\cite{Simon1998,Schmuedgen2017} for moments states that the positive definiteness of $\mathbf{S}$ is equivalent to the existence of some Hermitian operator $\hat H'$ that produces $M_n$ by $M_n=\lang\Phi|(\hat H')^n|\Phi\rang$, and {vice versa}. This is satisfied by some approximations, e.g. if the moments correspond to those of $\hat{P} \hat{H} \hat{P}$ where $\hat{P}$ is a linear projector onto a subspace (for example as in truncated configuration interation) but in general such a Hermitian operator may not exist for arbitrary many-body approximations to the moments.

In practice, therefore, we use the following termination criteria:
\begin{itemize}
    \item Power method (with Chebyshev acceleration). Terminate at the iteration $k$ where $E_k^\mathrm{P}(E_k^\mathrm{C})$ starts to increase. 
    
    \item Lanczos algorithm. Terminate when the (normalized) smallest singular value of $\mathbf{S}$ ($\sigma_\text{min}/\sigma_\text{max}$ where $\sigma$ denotes singular value) is below a positive threshold $\epsilon$. We use a threshold of $\epsilon = 10^{-10}$ in all our calculations. 
\end{itemize}

\section{Coupled Cluster Framework for Hamiltonian Moments}

We now describe a specific quantum many-body approximation for the Hamiltonian moments that is motivated by the well-studied coupled cluster theory\cite{Shavitt2009}. To simplify the discussion, we focus on the case where $|\Phi\rang = |\phi\rang$ is a single determinant. We start with the exact moment generating functions for single determinant expectation values, 
\begin{equation}
    M_\phi^X (\tau) = \langle \phi^X | e^{\tau\hat H} |\phi\rangle = \sum_n \frac{\tau^n}{n!} M_{\phi, n}^X
\end{equation}

where $|\phi^X\rang$ is shorthand for a determinant with excitation string $X$, and $M_{\phi,n}^X=\lang \phi^X|\hat H^n|\phi\rang$ are the transition Hamiltonian moments. When $\lang \phi^X|=\lang \phi|$, they reduce to the regular moments $M_{\phi,n}=\lang \phi|\hat H^n|\phi\rang$. We then write down a cluster factorization of $e^{\tau\hat H} |\phi\rangle$, 
\begin{equation}
    e^{\tau \hat H}|\phi\rang =  e^{S_{\phi}(\tau)+\hat W_{\phi}(\tau)}|\phi\rang
    \label{eqn:mCC-ansatz}
\end{equation}
in which $S_\phi(\tau)$ is a scalar and $\hat W_{\phi}(\tau)$ is expanded in cluster excitation operators:
\begin{equation}
    \hat W_{\phi}(\tau) = \sum_{m=1}^{\infty}\sum_{a_1\cdots a_m}\sum_{i_1\cdots i_m} [W_{\phi,m}(\tau)]_{a_1\cdots a_m}^{i_1\cdots i_m}\{a_1^\dagger i_1\cdots a_m^\dagger i_m\}
    \label{eqn:cgf}
\end{equation}
where $a_1\cdots a_m$ and $i_1\cdots i_m$ are virtual/occupied indices w.r.t. the reference determinant $|\phi\rang$ and $\{\cdots\}$ indicates normal-ordering w.r.t. $|\phi\rangle$. We note that $e^{\tau\hat H}|\phi\rang$ is the imaginary time evolved state (for time $-\tau$) and this cluster formulation is related to the one used in finite-temperature CC theory\cite{Sanyal1992,White2018}, although we here use the Schr{\"o}dinger picture rather than the interaction picture used in those works. 

The Taylor expansions of $S_\phi(\tau)$ and $W_{\phi,m}(\tau)$ around $\tau=0$ define the Hamiltonian cumulants: 
\begin{equation}
    S_\phi(\tau) = \sum_n \frac{\tau^n}{n!}\lang\phi|\hat H^n|\phi\rang_\mathrm{c} 
\end{equation}
\begin{equation}
    [W_{\phi,m}(\tau)]_{a_1\cdots a_m}^{i_1\cdots i_m} = \sum_n \frac{\tau^n}{n!}\lang\phi_{a_1\cdots a_m}^{i_1\cdots i_m}|\hat H^n|\phi\rang_\mathrm{c} \label{eq:transitioncumulants}
\end{equation}

By projecting (\ref{eqn:mCC-ansatz}) to suitable excited bras, we can verify that the Hamiltonian moments are related to the Hamiltonian cumulants by
\begin{equation}
    M_{\phi, n}^X = \sum_{m=1}^{n} \sum_{Y+Z=X} C_{n-1}^{m-1}  \lang \phi^Y|\hat H^m|\phi\rang_\mathrm{c} M_{\phi, n-m}^Z  \label{eqn:hcn2hn}
\end{equation}
where $\sum_{Y+Z=X}$ sums over all excitation strings $Y$ and $Z$ that add up to $X$. For $\lang \phi^X|=\lang \phi|$, equation (\ref{eqn:hcn2hn}) reduces to the well-known moment-cumulant relationship $M_{\phi, n} = \sum_{m=1}^{n}  C_{n-1}^{m-1}  \lang \phi|\hat H^m|\phi\rang_\mathrm{c} M_{\phi, n-m}$ in statistics, and (\ref{eqn:hcn2hn}) generalizes it to cases where $\lang \phi^X|\neq \lang \phi|$.

By taking the $\partial_\tau$ derivatives of (\ref{eqn:mCC-ansatz}) and projecting to suitable excited bras, $S_\phi(\tau)$ and $\hat W_\phi(\tau)$ satisfy the following differential equations
\begin{equation}
    \lang \phi|e^{-\hat W_\phi(\tau)}\hat He^{\hat W_\phi(\tau)}|\phi\rang = \frac{\partial }{\partial \tau} S_\phi(\tau)
\end{equation} 
\begin{equation}
    \lang \phi_{a_1\cdots a_m}^{i_1\cdots i_m}|e^{-\hat W_\phi(\tau)}\hat He^{\hat W_\phi(\tau)}|\phi\rang = \frac{\partial }{\partial \tau} [W_{\phi,m}(\tau)]_{a_1\cdots a_m}^{i_1\cdots i_m}
    \label{eqn:mCC-diff}
\end{equation} 

which can be solved formally by series expansion. Applying the Baker-Campbell-Hausdorff (BCH) formula to $e^{-\hat W_\phi(\tau)}\hat He^{\hat W_\phi(\tau)}$ and expanding both sides as Taylor series in $\tau$, we arrive at a set of equations which relate the higher order cumulants (r.h.s.) to the lower order cumulants (l.h.s.):
\begin{widetext}
\begin{equation}
    \lang\phi_{a_1\cdots a_m}^{i_1\cdots i_m}|\delta_{n0}\hat H+[\hat H, \hat W_{\phi}^{(n)}]_-+\frac{1}{2!}\sum_{m_1+m_2=n}\frac{n!}{m_1!m_2!}[[\hat H, \hat W_{\phi}^{(m_1)}]_-,\hat W^{(m_2)}_{\phi}]_-+\cdots |\phi\rang = \lang\phi_{a_1\cdots a_m}^{i_1\cdots i_m}|\hat H^{n+1}|\phi\rang_\mathrm{c}
    \label{eqn:mCC-discrete}
\end{equation}
\end{widetext}
where $\hat W_\phi^{(n)}$ is defined by
\begin{equation}
    \hat W_\phi(\tau) = \sum_{n}\frac{\tau^n}{n!}\hat W_\phi^{(n)}
\end{equation}
and serves as a shorthand notation for the terms produced when we substitute (\ref{eq:transitioncumulants}) into (\ref{eqn:cgf}). We note that just like in the conventional coupled-cluster theory, the BCH expansion terminates at the fourth order for  2-body Hamiltonians. 

The above iterative equations define a coupled-cluster framework for computing the Hamiltonian cumulants and thereby the Hamiltonian moments, and we refer to them as the moment CC (mCC) equations. By truncating the excitations in $\hat W_\phi(\tau)$ and/or the order of the nested commutators, we generate specific truncated coupled-cluster approximations to the moments, for example, coupled cluster singles and doubles approximate moments (mCCSD). It is worth mentioning that the doubly connected moments approximation proposed by Ganoe \textit{et al.} in Ref.~\onlinecite{Ganoe2023} corresponds to computing the moments with a truncation to linearized coupled-cluster doubles in this language, although the moments were not used in a linear algebra algorithm in that work. 

\begin{figure}[!ht]
    \centering
    \includegraphics[width=0.45\textwidth]{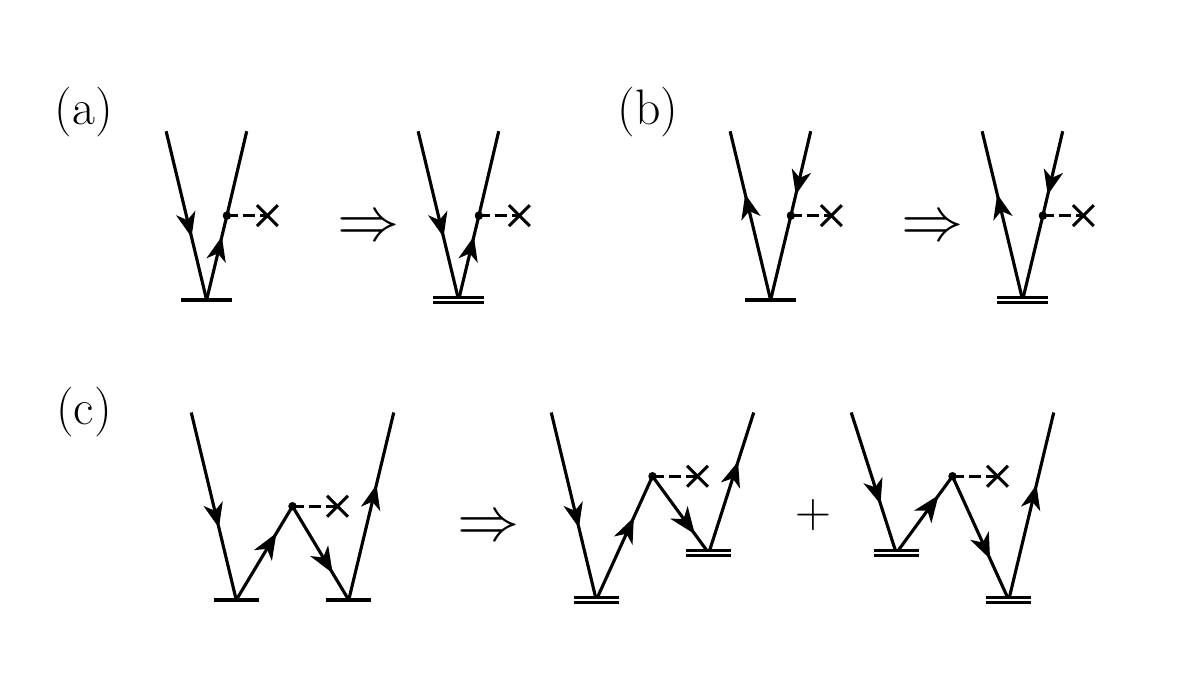}
    \caption{Coupled Cluster Singles (CCS) diagrams and their mCCS counterparts for a 1-body Hamiltonian. Single (double) lines correspond to CCS (mCCS) amplitudes $\hat T_1 (\hat W_1)$. When generating $\hat T_1$ by the standard CC iteration, energy denominators appear, but this is not the case for  $W_1$. Note also that for (c), translating from the CCS to mCCS diagrams introduces a combinatorial factor arising from the different time-orderings (here, the factor is 2, as seen by the two identical mCC diagrams in (c)).}
    \label{fig:ccs}
\end{figure}

The mCC equations (\ref{eqn:mCC-discrete}) can be expressed in terms of diagrams similar to those of the perturbation-order-resolved ordinary CC amplitude equation diagrams\cite{Shavitt2009}. The nested commutators in (\ref{eqn:mCC-discrete}) correspond to connected diagrams where $\hat H$ and $\hat W_\phi^{(n)}$ are represented by Hamiltonian/amplitude vertices. Each mCC diagram has an ordinary CC diagram counterpart with the same topology. The differences are that the amplitude vertices $\hat W_\phi^{(n)}$ are now free of energy denominators, and the mCC diagrams contain an extra multinomial coefficient, $n!/(m_1!\cdots  m_\alpha!)$ for the diagram containing vertices $\hat W_\phi^{(m_1)},\cdots,\hat W_\phi^{(m_\alpha)}$, as seen from (\ref{eqn:mCC-discrete}). This extra combinatorial factor serves to correct the Frantz-Mills diagram factorization theorem\cite{Frantz1960,Shavitt2009} as the original version no longer holds when the diagrams are generated without any energy denominators. Following this argument, given a subset of CC diagrams, one may obtain the corresponding mCC diagrams by properly assigning the extra coefficients (see Fig.~\ref{fig:ccs}). For example, although we do not focus on properties in this work, we can define mCC $\Lambda$ equations in this way. Similarly the relationship provides a translation between diagrammatic time-independent MBPT approximations and approximations to compute Hamiltonian moments.

\section{Quantum many-body linear algebra with coupled cluster Hamiltonian moments}

We now consider the use of coupled cluster Hamiltonian moments defined from the mCC equations in quantum many-body linear algebra. In the limit of no truncation, the mCC moments are exact, and the linear algebra algorithms in Sec.~\ref{sec:la} will all converge to the exact eigenvalue, similarly to CC without truncation. However, once truncation is introduced, the relationship between linear algebra using truncated mCC moments and the conventional truncated CC theory is unclear, because the solution conditions are different. For example, truncated CC solves a non-Hermitian eigenvalue problem, whereas the mCC Lanczos algorithm corresponds to  a Hermitian eigenvalue problem. Further, the termination criteria of the linear algebra algorithms do not correspond to the CC amplitude equations. We thus seek to understand the performance of the quantum many-body linear algebra via numerical examples below, computed using a preliminary implementation in the PySCF quantum chemistry package\cite{Sun2015,Sun2018,Sun2020}.

We first show the convergence behavior of the power method, Chebyshev accelerated power method, and Lanczos method, using mCCSD moments, for two illustrative systems, namely the H$_{10}$ ring in the minimal STO-6G basis~\cite{Hehre1969}  at two interatomic spacings $R_\mathrm{H-H}=1.03$~\AA~ (equilibrium) and $R_\mathrm{H-H}=1.95$~\AA~ (strongly correlated), and nitrogen dimer in the cc-pVDZ~\cite{Dunning1989} basis also at two spacings $R_\mathrm{N-N}=2.30a_0$ (equilibrium) and $R_\mathrm{N-N}=4.10a_0$ (stretched).
All moments were computed from restricted Hartree-Fock (RHF) determinants, and we show the results from restricted coupled cluster singles and doubles (RCCSD) and full configuration interaction (FCI) or near-exact density matrix renormalization group (DMRG)~\cite{zhai2021,zhai2023} for comparison. 

\begin{figure}[!ht]
    \centering
    \includegraphics[width=0.45\textwidth]{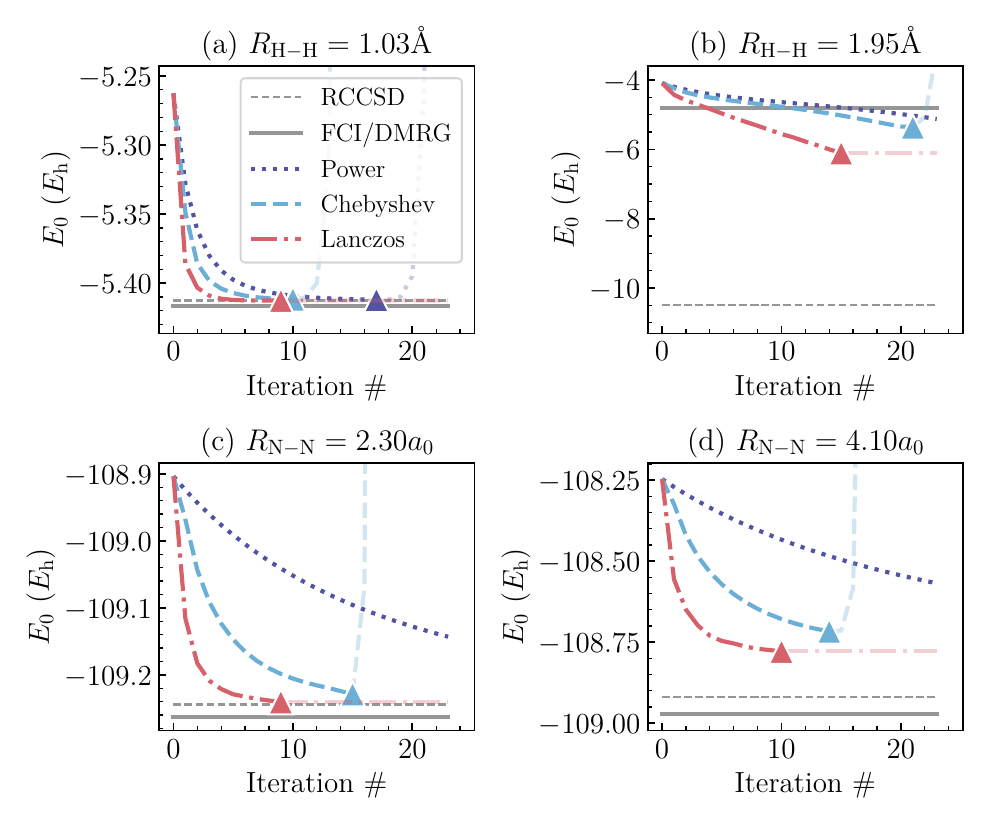}
    \caption{Convergence patterns of various quantum many-body linear algebra algorithms (power method, Chebyshev iteration, and Lanczos method) with approximate RmCCSD moments, compared to RCCSD and FCI(DMRG) on $\mathrm{H_{10}}(\mathrm{N_2})$. The symbols denote when the termination criterion has been reached.}
    \label{fig:H10_conv}
\end{figure}

In Fig.~\ref{fig:H10_conv}a we see that for the hydrogen ring at the equilibrium spacing, the mCCSD energies all converge smoothly as a function of iteration number, although at different rates, with Lanczos converging the fastest, then the Chebyshev iteration, and finally the power method. At the termination point of the iterations, the energies are already well converged, to within $\sim$0.1 m$E_\mathrm{h}$. In this system at this geometry, we expect CCSD to be very accurate and the mCCSD moments to be nearly exact, and thus we expect close agreement between the two. This is indeed the case, e.g. with the converged Lanczos mCCSD energy being $-5.412535\ E_\mathrm{h}$, compared to the converged CCSD energy of $-5.412538\ E_\mathrm{h}$.

In Fig.~\ref{fig:H10_conv}b, we see that for the hydrogen ring at the stretched geometry, the convergence is very different. Although the Lanczos overlap matrix is positive definite for a large number of iterations (up to $k=15$), the corresponding Lanczos-mCCSD energies become lower than that of FCI before the termination criterion is reached. A similar picture is seen using the (accelerated) power method, which terminates at a somewhat higher energy than with the Lanczos iteration. Both results are also quite different from those obtained using CCSD, which gives an even more non-variational energy, from which we conclude that the linear algebra serves to reduce the overestimation of the correlation energy.
A further difference is that the conventional CCSD energy is extremely hard to converge because the amplitude equations become hard to solve when the amplitudes are large. However, there is no such difficulty in evaluating the moment equations in (\ref{eqn:mCC-discrete}), which do not require any type of nonlinear solver.

\begin{figure}[!ht]
    \centering
    \includegraphics[width=0.45\textwidth]{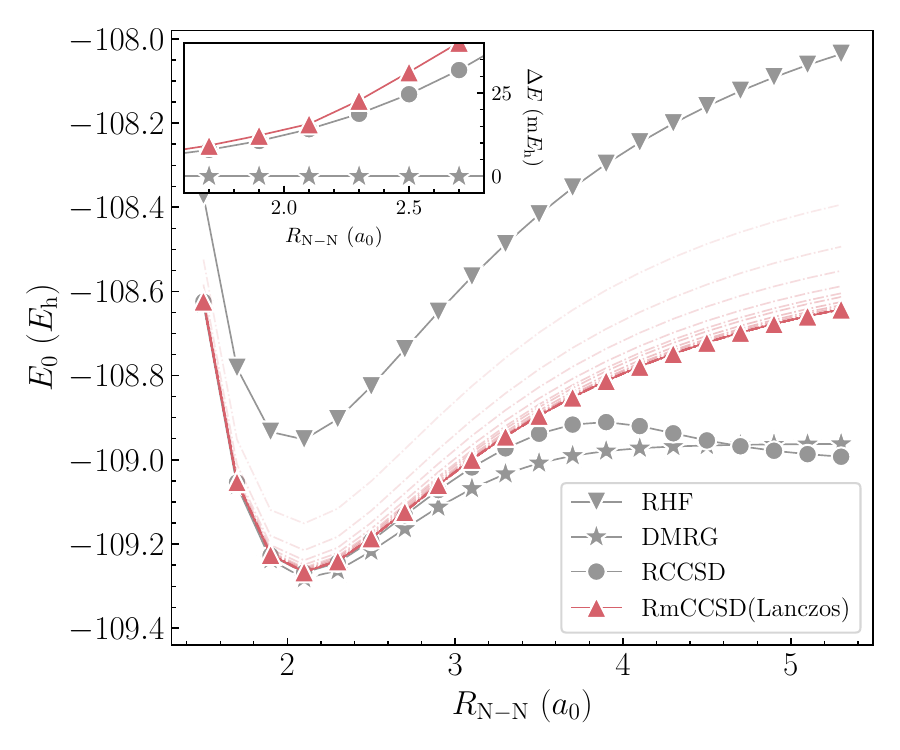}
    \includegraphics[width=0.45\textwidth]{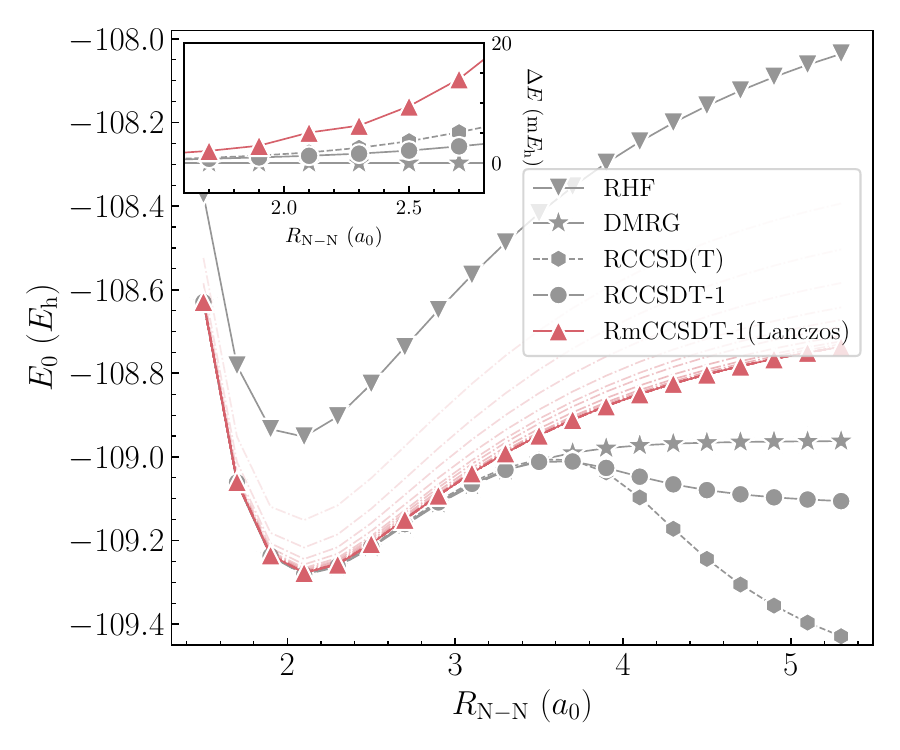}
    \caption{Potential energy curves of $\rm N_{2}$ obtained by the quantum many-body Lanczos algorithm with RmCC moments truncating to singles and doubles (RmCCSD, upper panel), and an approximate triples (RmCCSDT-1, lower panel). The dashed-and-dotted traces from top to bottom show Lanczos convergence with increasing iterations $k$.}
    \label{fig:N2_RmCCSD}
\end{figure}

\begin{figure}[!ht]
    \centering
    \includegraphics[width=0.45\textwidth]{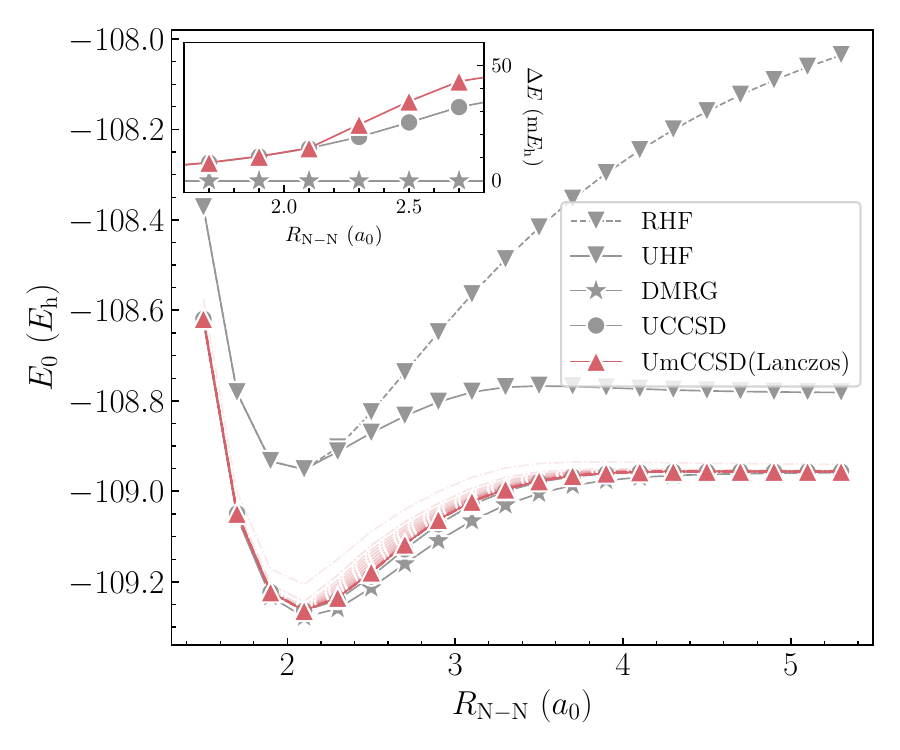}
    \includegraphics[width=0.45\textwidth]{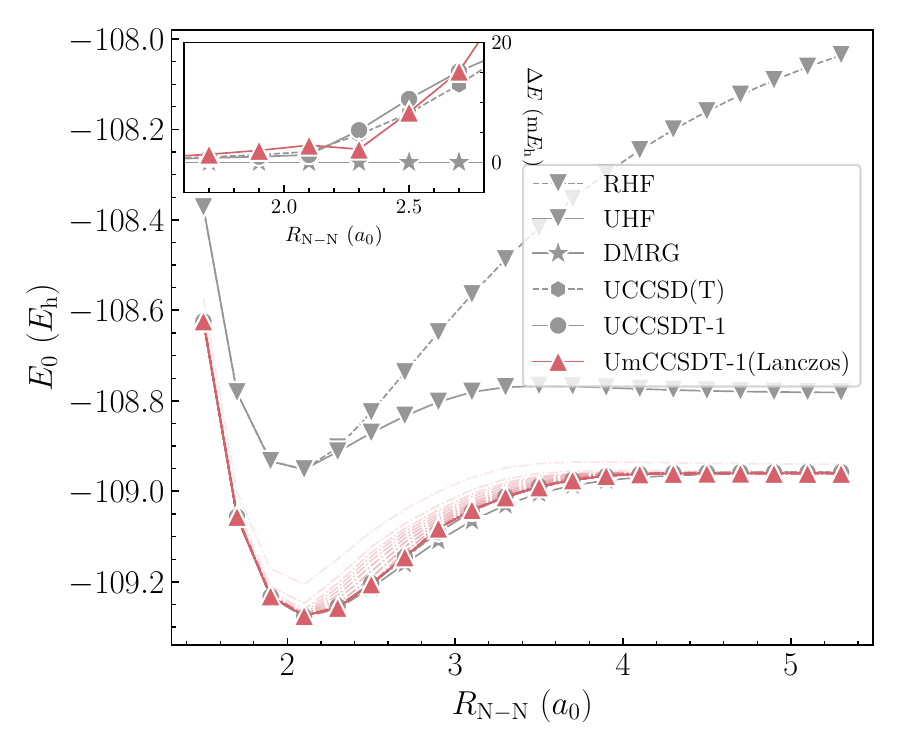}
    \caption{Potential energy curves of $\rm N_{2}$ obtained by the quantum many-body Lanczos algorithm with UmCC moments truncating to singles and doubles (UmCCSD, upper panel), and an approximate triples (UmCCSDT-1, lower panel). The dashed-and-dotted traces from top to bottom show Lanczos convergence with increasing iterations $k$.}
    \label{fig:N2_UmCCSD}
\end{figure}

\begin{figure}[!ht]
    \centering
    \includegraphics[width=0.45\textwidth]{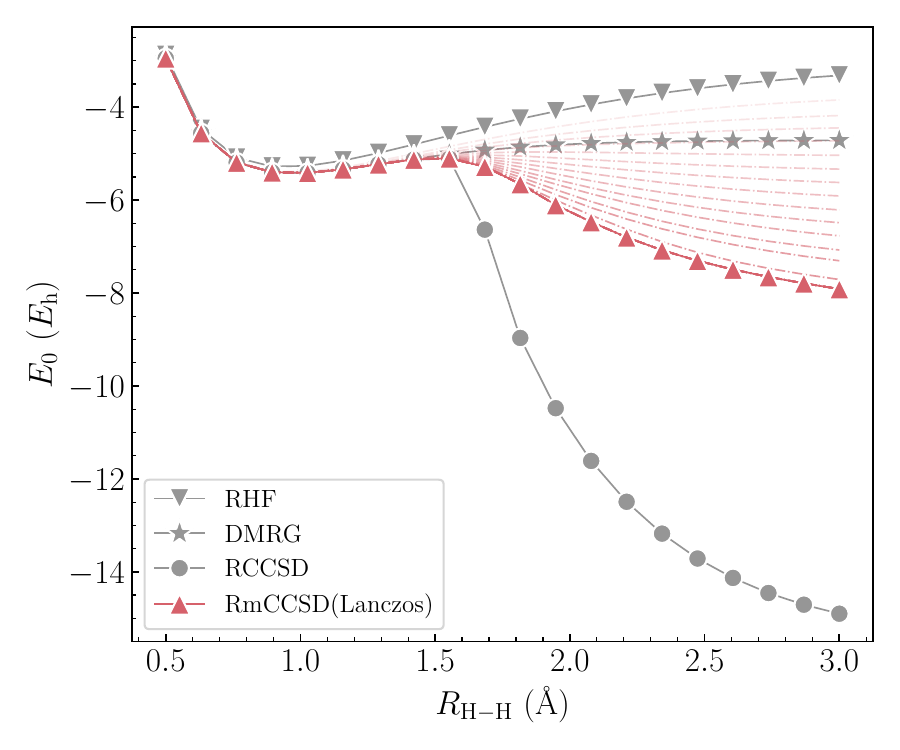}
    \caption{Potential energy curves of $\rm H_{10}$ ring obtained by the quantum many-body Lanczos algorithm with RmCC moments truncating to singles and doubles (RmCCSD). The dashed-and-dotted traces from top to bottom show Lanczos convergence with increasing iterations $k$.}
    \label{fig:H10_mCCSD}
\end{figure}

In Figs.~\ref{fig:N2_RmCCSD}-\ref{fig:H10_mCCSD}, we show the full potential energy curve of the nitrogen dimer and the $\rm H_{10}$ ring
focusing on results from the Lanczos algorithm. In addition to using the mCCSD moments, we also show results from mCCSDT-1 moments (defined by analogy with the CCSDT-1\cite{Lee1984,Lee1984a} truncation, which is an iterative approximate triples correction with $O(N^7)$ scaling).
Results from standard CCSD, CCSD(T) (CCSD with perturbative triples\cite{Raghavachari1989}), and CCSDT-1 are shown as well.

From Fig.~\ref{fig:N2_RmCCSD} we see in the nitrogen dimer that Lanczos-mCCSD and mCCSDT-1 (starting from a RHF reference) again perform similarly to CCSD and CCSDT-1 in the equilibrium region. However as the bond is stretched, the Lanczos-mCC results significantly improve on the standard truncated CC energies. It is well known that CCSD has a turnover and becomes non-variational at longer bond-distances in this system, contributing to a severe breakdown of CCSD(T) and, to a lesser degree, CCSDT-1. However,  the Lanczos-mCCSD and Lanczos-mCCSDT-1 results remain variational across the potential energy curve and the energies at long distances appear as a smooth continuation of the results at shorter distances. Also, as noted above, while CCSD becomes quite hard to converge for the longer bond-lengths, the Lanczos-mCC calculations do not suffer from this problem.
Fig.~\ref{fig:N2_UmCCSD} shows the same results starting from an unrestricted Hartree-Fock reference. In this case, the accuracy around equilibrium and at longer geometries is similar between the methods, with the Lanczos-mCCSD displaying slightly larger errors than UCCSD in the spin-recoupling region, with the Lanczos-mCCSDT-1 behaving similarly to CCSD(T) and CCSDT-1.

The full potential energy curve of the hydrogen ring in Fig.~\ref{fig:H10_mCCSD} further illustrates the points seen already at the two geometries chosen previously. Around the equilibrium geometry, the Lanczos-mCC results are visually indistinguishable from their truncated CC counterparts. At long bond lengths, although the Lanczos-mCC results greatly improve on the standard CC overestimation of the energy, there is still a similar turnover point where they start to become inaccurate. 

\section{Outlook}

We conclude from our numerical experiments that the translation of diagrammatic theories into approximations for linear algebra primitives, specifically Hamiltonian moments, gives rise to a new class of approximate quantum many-body methods. Intriguingly, we find that these can behave quite differently from the many-body perturbation framework from which they are inspired. For example, using coupled-cluster inspired moments in a Lanczos algorithm avoids the difficult solution of the non-linear amplitude equations and, in some cases, avoids the variational collapse of the standard coupled cluster theory. While we have focused on diagrammatic analogies to coupled cluster theory in this work, generalizations to other types of many-body approximations, including those based on Green's functions, or which start from multi-reference states, is clearly possible, and desirable in the future.

\section{Acknowledgements} This work was supported by the US Department of Energy, Office of Science, Basic Energy Sciences, via grant no. DE-SC0018140. GKC is a Simons Investigator in Physics.

\bibliography{mcc.bib}

\end{document}